\documentclass[aps,prl,twocolumn,showpacs]{revtex4}

\usepackage{amssymb,amsmath,graphicx,graphics,color}

\newcommand{\de}{\partial}

\newcommand{\eq}[2]{\begin{equation} \label{#1} #2 \end{equation}}

\newcommand{\etal}{{\em et al.}}

\begin{document}

\title{Super-resonant radiation stimulated by high-harmonics}
\author{Cristian Redondo Lour\'es}
\author{Thomas Roger}
\author{Daniele Faccio}
\author{Fabio Biancalana}
\affiliation{School of Engineering and Physical Sciences, Heriot-Watt University, EH14 4AS Edinburgh, UK}

%\begin{figure}
%\includegraphics[width=7cm]{fig1.jpg}
%\caption{(Color online) }
%\label{fig1}
%\end{figure}

\begin{abstract}
Solitons propagating in media with higher order dispersion will shed radiation known as dispersive wave or resonant radiation, with applications in frequency broadening, deep UV sources for spectroscopy or simply fundamental studies of soliton physics. Starting from a recently proposed equation that models the behaviour of ultrashort optical pulses in nonlinear materials using the analytic signal, we find that the resonant radiation associated with the third-harmonic generation term of the equation  is parametrically stimulated with an unprecedented gain. Resonant radiation levels, typically only a small fraction of the soliton, are now as intense as the soliton itself. The mechanism is quite universal and works also in normal dispersion and with harmonics higher than the third. We report experimental hints of this super-resonant radiation stimulated by the fifth harmonic in diamond.
\end{abstract}

\pacs{42.65.-k Nonlinear optics; 42.65.Ky Frequency conversion; harmonic generation, including higher-order harmonic generation; 42.65.Sf Dynamics of nonlinear optical systems; optical instabilities, optical chaos and complexity, and optical spatio-temporal dynamics; 42.65.Wi Nonlinear waveguides}
\maketitle

\paragraph{Introduction ---}

The process of resonant radiation emission in nonlinear media is extremely general and has been studied in a plethora of different systems, like solitons in fibers and bulk media \cite{elgin,Akhmediev,Husakou,biancalanaskryabin,skryabinluan,dudley}, three dimensional light bullets \cite{durand1,durand2,roger}, dispersive shock waves \cite{trillo1}, passive resonators \cite{res1,res2,res3,res4,res5}, and even more complex scenarios combining a mixture of the above \cite {trillo2}.
The shedding of this light is dictated by a nonlinear momentum conservation, i.e. the requirement is that the momentum of the pump is equal to the linear momentum of the dispersive wave propagating in normal dispersion \cite{Akhmediev,biancalanaskryabin,skryabinluan}. The condition of the momenta being equal is known as the \emph{phase matching condition}. In particular, for a system governed by the nonlinear Schr\"odinger equation (NLSE) it can only occur when a third (or higher) order dispersion term is present. In a recent  work, Conforti \emph{et al} proposed an equation for the {\em analytic signal} of an electric field that is formally similar to the NLSE but does not suffer from many of the limitations of the latter \cite{conforti} and only relies the reasonable assumption of neglecting backward propagating waves \cite{kinsler}. This equation has been found to predict some features of the nonlinear interaction between light and matter that were not present in the original NLSE, related to the so-called {\em negative frequency components} of the electromagnetic pulse \cite{amir,rubino,spm,cavity,mclenaghan}. In their paper, the authors discuss new phase matching conditions that arise from the new nonlinear polarisation terms in their equation, and theoretically predict the emission of what has been dubbed  \emph{negative resonant radiation} (NRR)  and \emph{third-harmonic resonant radiation} (THRR). The former had been previously identified experimentally by Rubino \emph{et al.} in their landmark experiment \cite{rubino,articlefabio}. However, the new THRR term was located in the deep infrared region of the spectrum for the physical system and material analysed (fused silica), where it was not efficiently fed by the pump and therefore has never been observed experimentally or numerically.

In the present paper, we explore the possibility of promoting the THRR signal into a very strongly resonant mode. We have found that when the THRR frequency is close to a high-harmonic frequency, a surprisingly large amount of the pump energy can be transferred to the radiation via a {\em stimulated} process. This is a two step mechanism: the pump propagating through the material releases energy to a higher harmonic, and the high-harmonic energy is then  transferred to the resonant THRR mode. This mode then appears as a very sharp, intense peak in the final output spectrum. The surprising defining properties of this novel radiation, which we dub \emph{super-resonant radiation} (SRR), is its extremely powerful gain dynamics and the unprecedented transfer of energy that occurs from the soliton to the radiation itself. These properties set the SRR apart from any currently known dispersive wave emission from solitons, with extremely interesting potential uses in frequency conversion applications. In the final part of the paper, we show a clear experimental hint of SRR in diamond, where intense pulses in normal dispersion are used to excite the THRR, which is then promoted to SRR when its frequency is close to the {\em fifth} harmonic of the pump.

\paragraph{Governing equations ---}
The equation proposed by Conforti \emph{et al} \cite{conforti} is, in dimensionless units,
\begin{widetext}
\eq{eqconforti}
{i\partial_\xi A+\hat{D}(i\partial_\tau)A+\left(1+\frac{i}{\mu}\partial_{\tau}\right) \bigg[  |A|^2A+ |A|^2A^\ast\exp(2i\phi)+\frac{1}{3}A^3\exp(-2i\phi)\bigg] _+=0 }
\end{widetext}
where $A=A(\xi,\tau)$ is the envelope of the analytic signal of the electric field, $\xi$ and $\tau$ are the dimensionless space and time variables (scaled with the second order dispersion length $L_{\rm D}\equiv t_{0}^2/|\beta_{2}|$ and the input pulse duration $t_{0}$, respectively), $\hat{D}\equiv\sum_{m=2}^{\infty}b_{m}(i\de_{\tau})^{m}/m!$ is the dispersion operator, $b_m=\beta_m/(|\beta_2|t_0^{m-2})$ are the normalised dispersion coefficients, $\phi\equiv \kappa\xi+\mu\tau$, $\kappa\equiv (\beta_1\omega_0-\beta_0)L_{\rm D}$ is a crucial parameter that measures the difference between the group and the phase velocities, $\mu=\omega_0t_0$ is the normalised pulse frequency, $\beta_1$ is the inverse group velocity, $\beta_0/\omega_0$ the inverse phase velocity, $L_{\rm D}\equiv t_{0}^{2}/|\beta_{2}|$ is the dispersion length, and $\omega_0$ the central frequency of the pulse. Equation (\ref{eqconforti}) has been successfully applied to optical fibers, crystals \cite{conforti,spm} and fiber or microring cavities \cite{cavity}.
 
Since by definition the {\em analytic signal} $\mathcal{E}=A\exp(i\beta_0z-i\omega_0t)$ is the positive frequency part of the electric field $E$, one has that $E$ can be written as $E=(\mathcal{E}+\mathcal{E}^\ast)/2$, where, by adding a field with only positive frequencies ($\mathcal{E}$) and another one with only negative components ($\mathcal{E}^\ast$) we obtain a field with a symmetrical Fourier spectrum (i.e., a \emph{real} field, $E$) \cite{amir}. In the absence of nonlinear interaction the evolution equations for $\mathcal{E}$ and $\mathcal{E}^\ast$ are completely decoupled, but the nonlinear polarisation in Eq. (\ref{eqconforti}) mixes both fields in a non-trivial way. The first term of the polarisation inside the square brackets in Eq. (\ref{eqconforti}) corresponds to the usual Kerr term. The third term is the well-known third-harmonic generation, and the second is the so called \emph{negative-frequency Kerr term} \cite{conforti}. The subscript symbol $+$ in the nonlinear part of Eq. (\ref{eqconforti}) means that spectral filtering must be performed, since $A$ must contain only the positive frequency components \cite{conforti,spm,cavity,stein1,stein2}.  

\begin{figure}
\includegraphics[width=8cm]{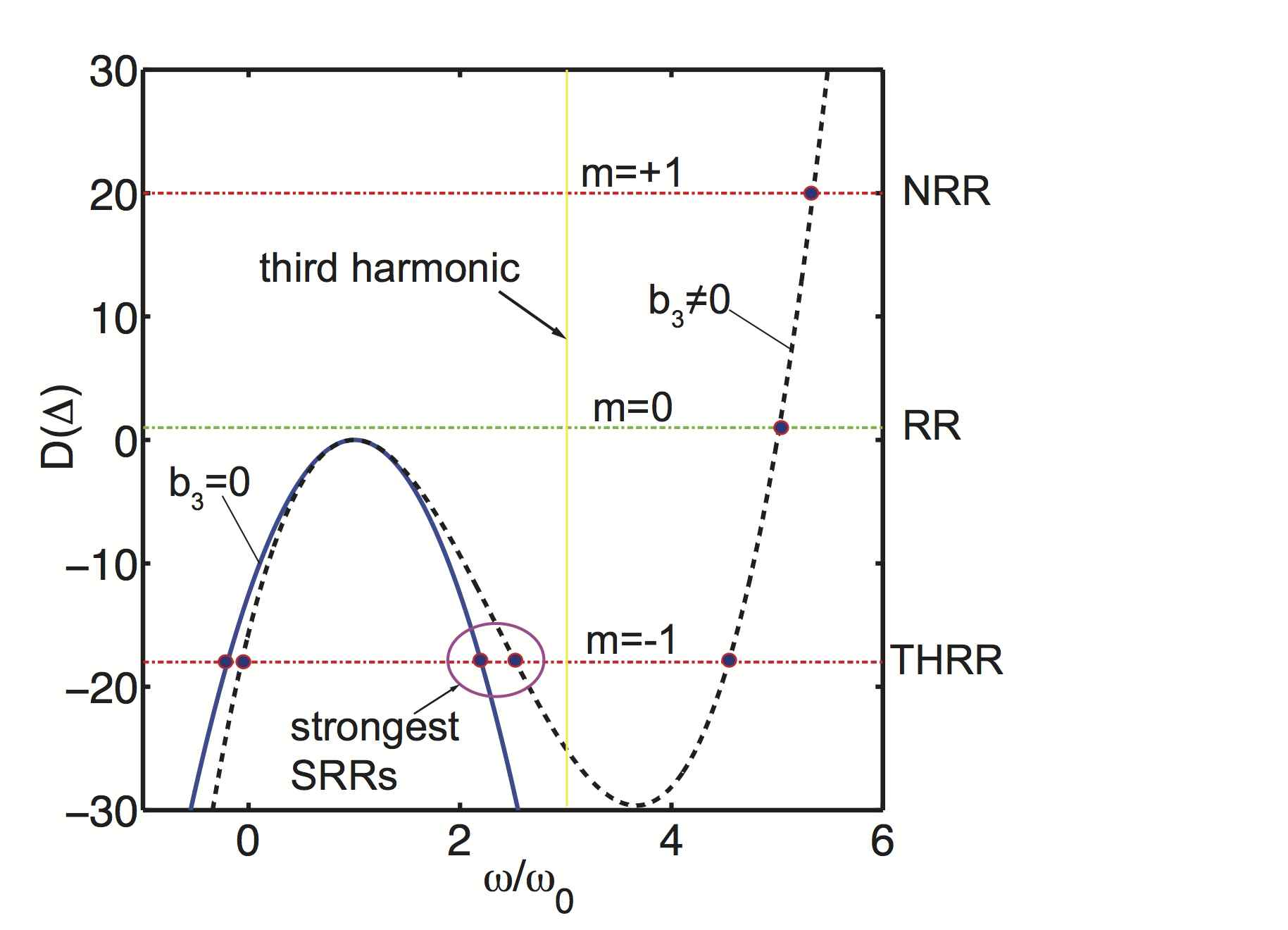}
\caption{(Color online) Phase matching curves from Eq. (\ref{phasematch}). The three horizontal lines represent the three phase matching conditions: NRR (upper, red line), RR (medium, green line) and THRR (lower, brown line). The two curves represent the dispersion for $b_2$ only (blue, solid line) and $b_3=0.15$ (black, dashed line). $\mu=5$ and $\kappa=10$ in both cases.}
\label{fig1}
\end{figure}

In Ref. \cite{conforti} all the phase matching conditions for the emission of resonant radiations by solitons have been derived -- there are {\em three} in total, one associated with each term of the nonlinear polarisation. These phase matching conditions can be written as:
\eq{phasematch}
{D(\Delta )=2m\kappa-(2m-1)q,}
where $q=P/2$, with $P$ the normalised power of the incident pulse (our equations are scaled so that $P=1$ always, so $q=1/2$), and $m=1$ for NRR, $m=0$ for the usual RR, and $m=-1$ for the THRR, see also Ref. \cite{conforti}. $D(\Delta)=\sum_{n=2}^\infty b_n\Delta^n/n!$ is the Fourier transform of the dispersion operator, where $\Delta$ is the dimensionless detuning between pulse and radiation. Note that, since $q>0$ and, in experimentally accessible conditions, $\kappa\gg q$, if we are in deep anomalous dispersion ($b_2<0$ and all other dispersion coefficients can be ignored) $D(\Delta)\equiv b_{2}\Delta^{2}/2\leq0$ and neither the phase matching for RR nor the one for NRR can be satisfied, see blue solid curve in Fig. \ref{fig1}. However, the phase matching for THRR can be fulfilled for two values of the detuning, one positive and one negative (see blue solid curve and dots showing crossings in Fig. \ref{fig1}). In the same figure we can see that when we include $b_3$, all three phase matching conditions can be satisfied for values of $\Delta>0$, and there are three different detunings for which we expect to find THRR, see dashed black curve and dots showing the crossings in Fig. \ref{fig1}.

In our numerical simulations, shown in the following, we have found that, when the position of the THRR is close to the third-harmonic that is created by the pump as it propagates through the medium, the radiation will grow rapidly and appears as a narrow, very intense peak in the spectrum in the position predicted by the phase matching condition (\ref{phasematch}), with $m=-1$. At variance with previously known dispersive wave emissions in fibers or bulk, this is a {\em two step mechanism}: the pulse gives energy to its third-harmonic during propagation, and then most of this energy is transferred to the phase-matched THRR closest to the third-harmonic frequency. This last step can only occur if the THRR is spectrally located close to the third harmonic frequency (see purple oval in Fig. \ref{fig1}). We therefore say that the THRR has been `promoted' to SRR. The surprising fact is that this effect is extremely efficient: the third-harmonic never manages to fully grow,since the THRR continuously absorbs almost all its energy, leading to the formation of an extremely intense and spectrally well-localized SRR peak.

\paragraph{Numerical simulations ---} 

\begin{figure}
\includegraphics[width=8cm]{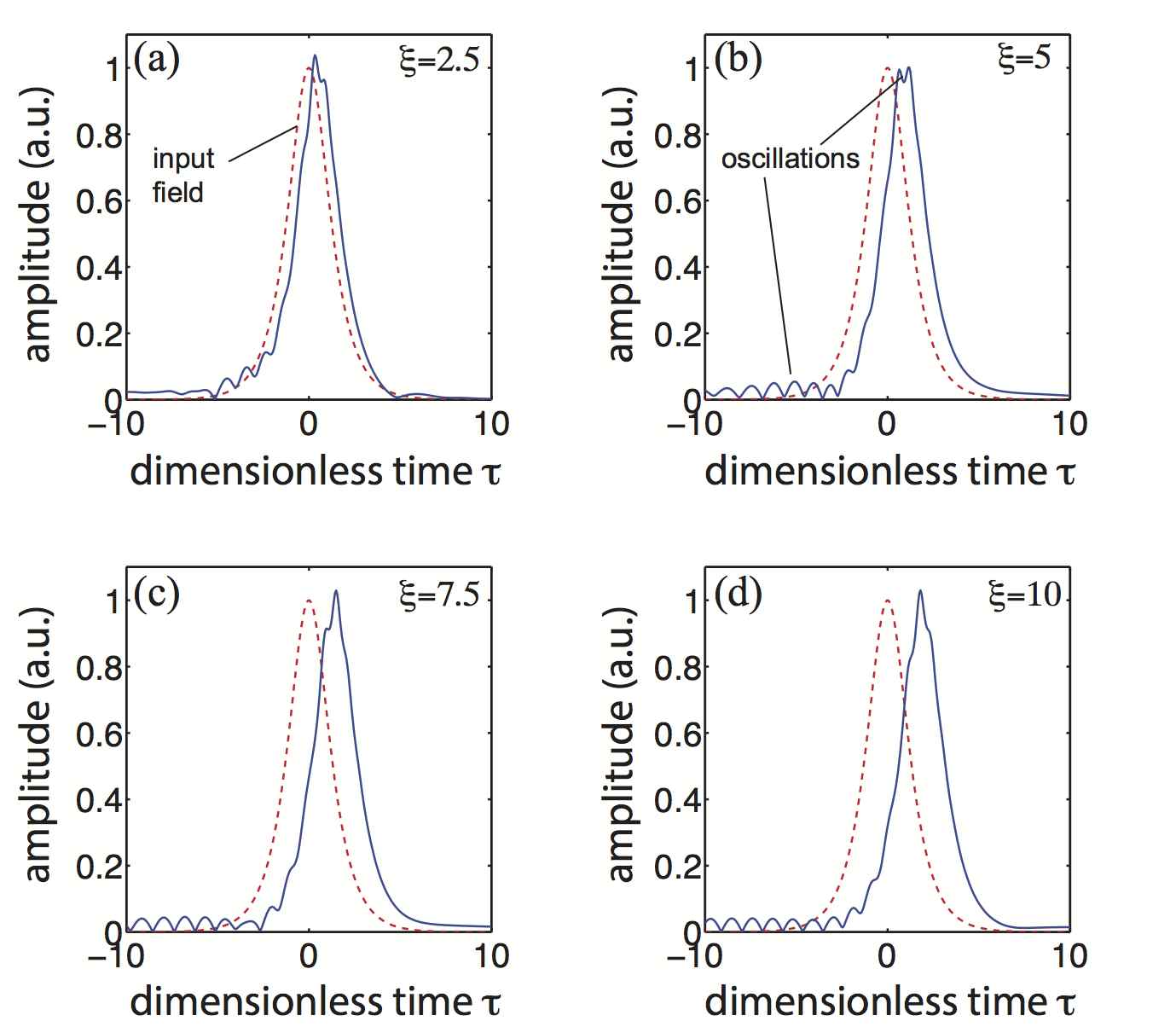}
\caption{(Color online) Soliton in time domain after a propagation of $\xi=2.5$ (a), $5$ (b), $7.5$ (c) and $10$ (d). We can see the oscillations on top of the pulse that leave it through the leading edge. Parameters: $b_3=0$, $\mu=5$ and $\kappa=10$. See also animation 1 in the Supplementary Material.}
\label{beta2_time}
\end{figure}

Figure \ref{beta2_time} shows the evolution in the time domain of an input hyperbolic secant pulse with $b_2<0$, $b_3=0$, $\mu=5$ and $\kappa=10$ after propagating $\xi=10$ dispersion lengths. For illustration purposes, these parameters are chosen in such a way that the THRR is phase-matched at a frequency between the pump ($\omega/\omega_{0}=1$) and its third-harmonic ($\omega/\omega_{0}=3$), around $\omega/\omega_{0}\sim 2$. We can observe how an oscillation appears on the top of the pulse in the time domain and then moves faster than the soliton, thus creating a leading oscillating tail. These violent intra-soliton oscillations are very characteristic of the SRR.

\begin{figure}
\includegraphics[width=8cm]{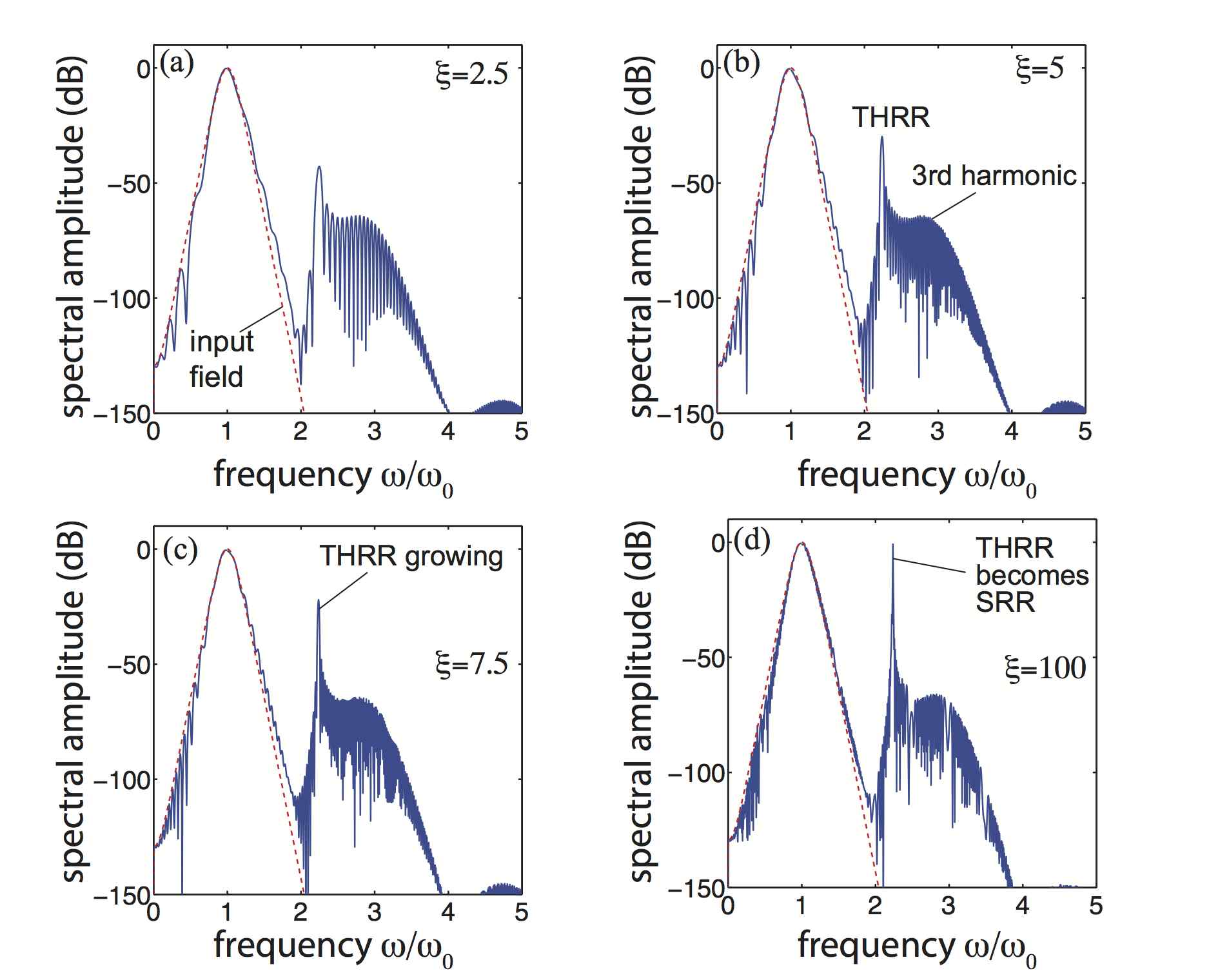}
\caption{(Color online) Initial (red dashed) and final (thick blue) spectrum of a pulse after a propagation of $\xi=2.5$ (a), $5$ (b), $7.5$ (c) and $100$ (d), with other parameters as in figure \ref{beta2_time}. We can clearly see here the evolution of a THRR peak into a stimulated SRR, at exactly the position predicted by the phase matching conditions (\ref{phasematch}) for $m=-1$ (vertical black dotted line). See also the intersection between the blue solid line and the THRR horizontal line in the purple oval of Fig. \ref{fig1}. See also animation 2 in the Supplementary Material.}
\label{beta2_spec}
\end{figure}

The evolution of the spectrum of this pulse is shown in Fig. \ref{beta2_spec}. We can see that the spectrum develops a very intense peak at exactly the position predicted for the THRR, see Eq. (\ref{phasematch}). This peak starts as a small peak in the third-harmonic peak but keeps growing with propagation, as energy is sucked from the third-harmonic. Note that for $\xi=100$ this peak has grown to be more intense than the pump pulse, since the THRR is oscillating on the top of the pulse itself. This is the situation in which the THRR is parametrically stimulated by the third-harmonic, and is promoted to SRR.

\begin{figure}
\includegraphics[width=8cm]{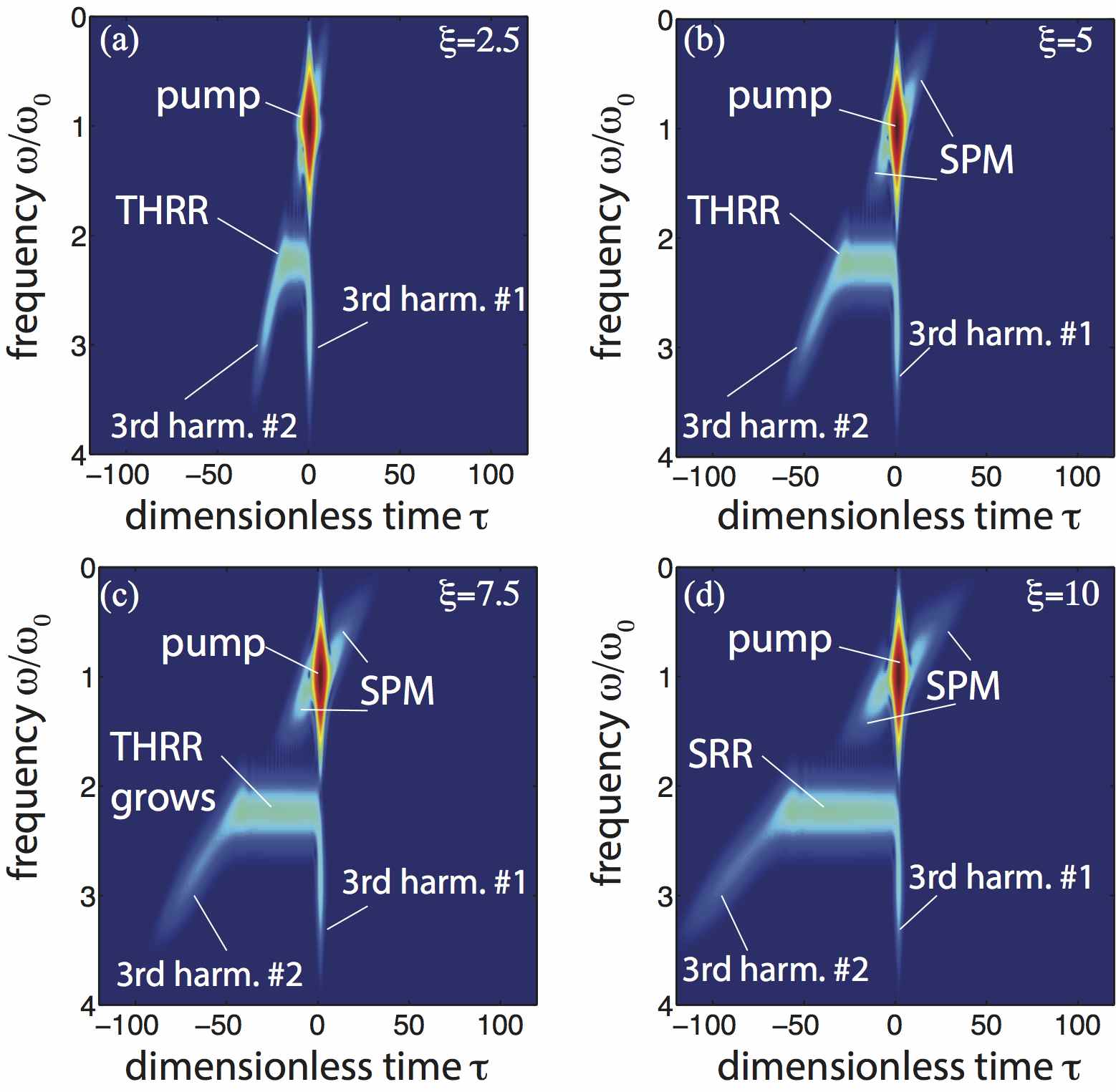}
\caption{(Color online) XFROG spectrograms for propagation lengths of $\xi=2.5$ (a), $5$ (b), $7.5$ (c) and $10$ (d). We can see the two components of the third-harmonic described in the text and the SRR between them. Note how the SRR band grows indefinitely during the $\xi$-evolution. See also animation 3 in the Supplementary Material.}
\label{beta2_xfrog}
\end{figure}

In Fig. \ref{beta2_xfrog} we show the XFROG spectrograms of the pulse evolution for $\xi=2.5$, $5$, $7.5$ and $10$, again for the case $b_{3}=0$. We see that the third-harmonic radiation has two components, one that propagates alongside the pulse and another one leading it (labeled `$\#$1' and `$\#$2', respectively). The SRR extends between these two components of the third-harmonic, which further confirms our hypothesis that SRR is indeed THRR stimulated by the higher harmonics inside the nonlinear material. Conventional self-phase modulation (SPM) lobes also appears close to the pump.

\begin{figure}
\includegraphics[width=8cm]{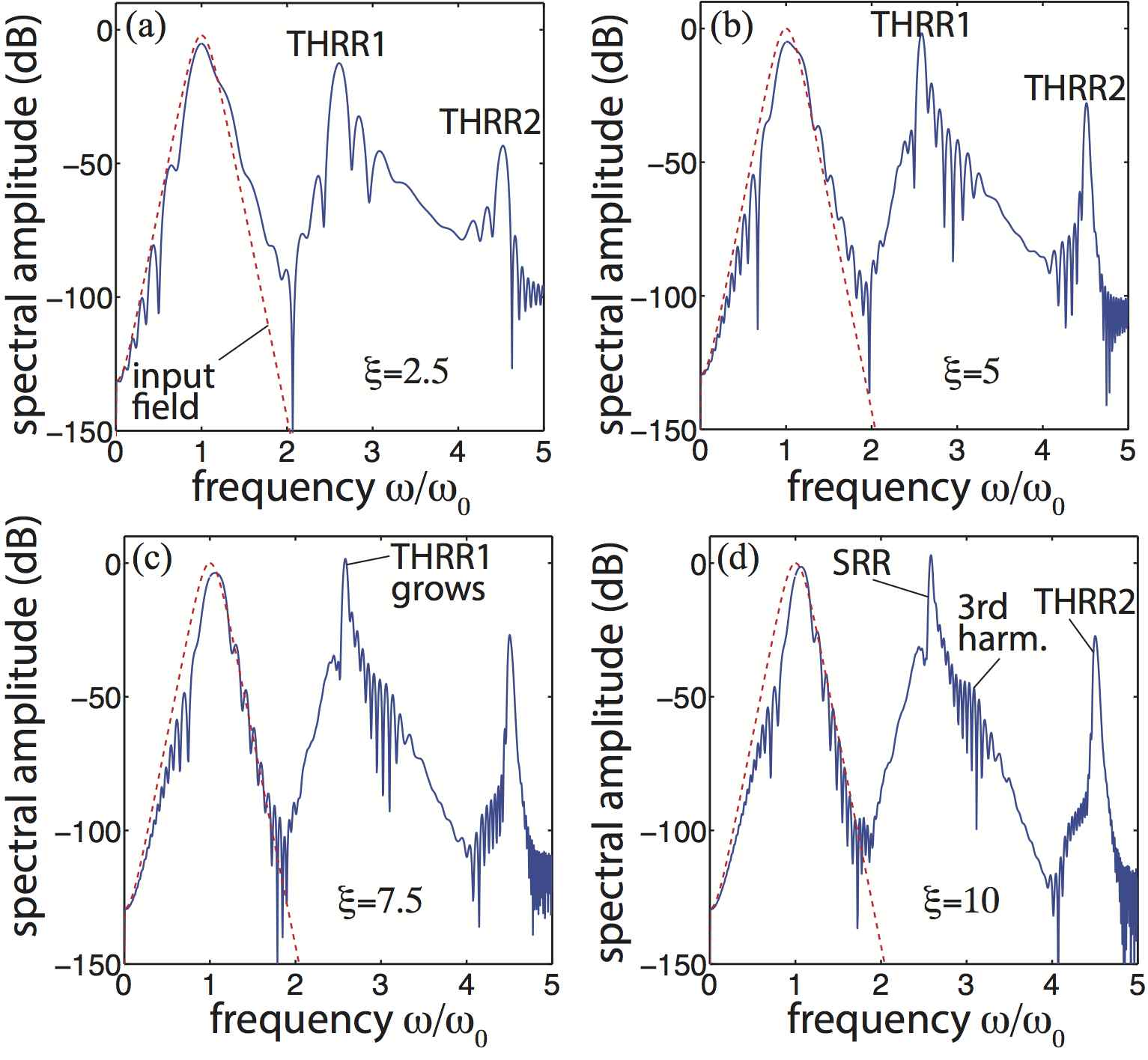}
\caption{(Color online) Snapshots of the initial (red dashed line) and final (blue solid line) spectra during a propagation of $\xi=10$ in the case $b_3=0.15$. The two vertical lines show the predicted position of the two THRRs. The THRR closer to the third-harmonic peak grows taller than the pump for $\xi=10$, and thus becomes a SRR. Other parameters are as in Fig. \ref{beta2_time}.}
\label{beta3_spec}
\end{figure}

In the case where $b_3 = 0.15$, with the same values of $\kappa$ and $\mu$ used before, the situation changes significantly. As seen in Fig. \ref{fig1}, Eq. (\ref{phasematch}) predicts two phase-matched frequencies near the third-harmonic (see the intersection between the black dashed line and the horizontal THRR line). Both of these frequencies are in a region of normal dispersion but due to the third-order dispersion term, one will be slower than the soliton and the other one will be faster. This means that in the time domain there will be two radiations, one of them leading the pulse and the other one trailing it. 
The output spectra for the case $b_{3}\neq0$ are shown in Fig. \ref{beta3_spec}. We can see that two peaks appear near the positions predicted by the phase matching conditions, with the one closer to the third-harmonic frequency growing much faster than the other. Again, the radiation peak closer to the third-harmonic (which is inside the purple oval in Fig. \ref{fig1}) is a stimulated THRR which is then promoted to SRR, consistently with the theoretical interpretation given above.

\paragraph{Experimental hints of SRR ---} In order to support the previous theoretical considerations, we show here experimental evidence of THRR stimulated by the 5th-harmonic in diamond. Odd harmonics higher than the 3rd are generated in the sample during propagation due to cascaded four-wave mixing, for which $\chi^{(3)}$ is responsible. Therefore the same stimulated SRR could in principle appear when the THRR frequency is close to the 5th-harmonic rather than the 3rd-harmonic, but with a much smaller amplitude. The use of the 5th-harmonic instead of the 3rd is at times useful in some materials, due to the very unclean spectra surrounding the 3rd-harmonic when pumping with very high energies.

For our proof-of-concept, we have used $50$ fs pulses injected in a $500$ $\mu$m bulk diamond.  An amplified Ti:Sapphire laser with central wavelength $\lambda_0 = 785$ nm is used to pump an optical parametric amplifier (OPA, TOPAS-C, Light Conversion Ltd.) producing infrared light pulses whose wavelength can be tuned between $1750$ and $2050$ nm. The resulting pulses are produced at a repetition rate of $100$ Hz and with pulse duration $70$ fs. The IR pulses are focused with an $f = 150$ mm lens to a spot radius of $\sim 36 \mu$m providing a peak  intensity of $I = 28$ TW/cm$^2$. A  single crystal diamond cut along the $\langle100\rangle$ axis is used to study the dynamics of the THRR as the pump wavelength is tuned. The output of the diamond crystal is imaged onto a spectrometer (Andor Shamrock 303i and iDus DU420A) providing visible spectrum data. In order to isolate the 5th harmonic from the intense 3rd harmonic contribution and have enough dynamic range, the high frequency component ($\lambda < 510$ nm) is blocked inside the spectrometer.

\begin{figure}
\includegraphics[width=8cm]{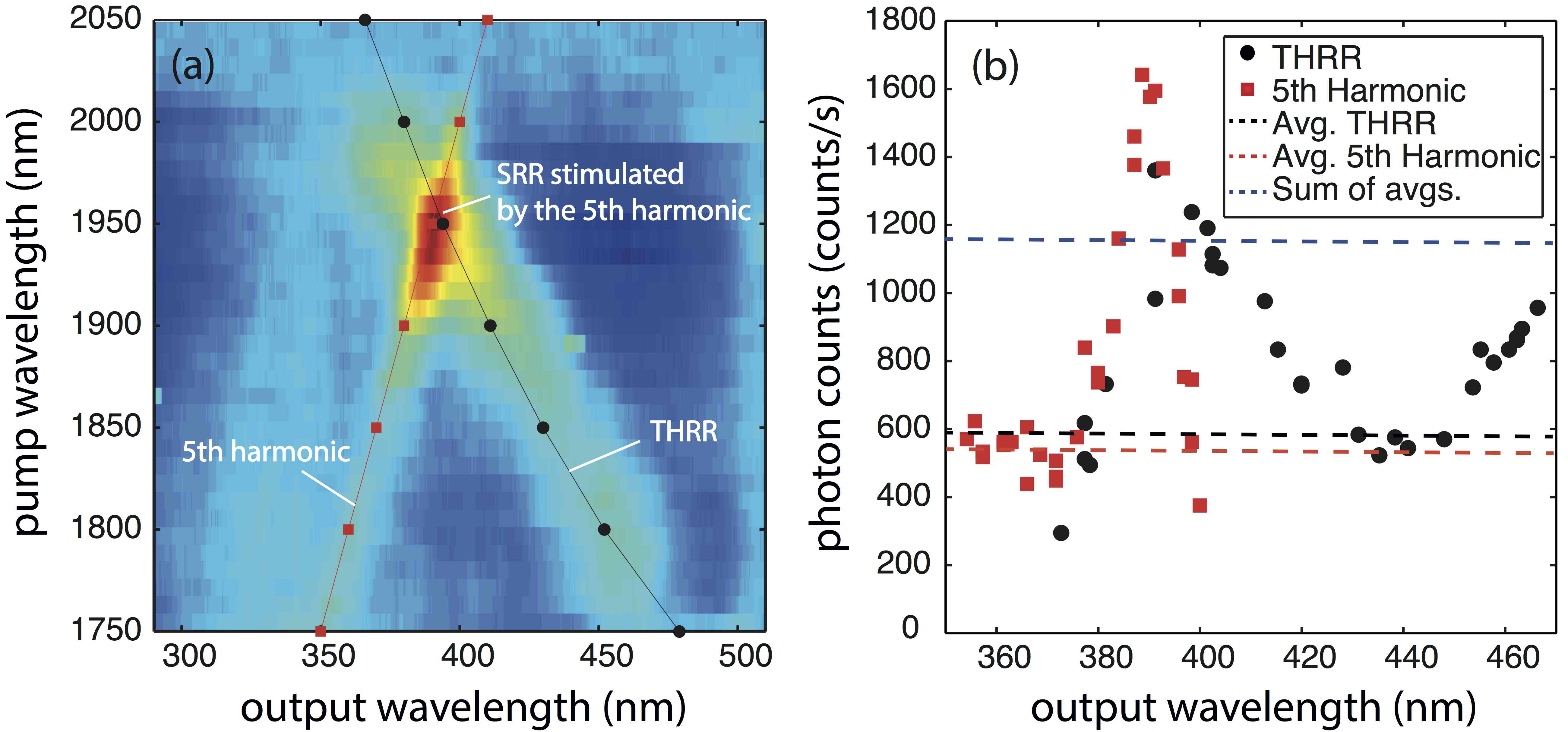}
\caption{(Color online) (a) Emission of THRR (black dots) and 5th harmonic (red squares) in diamond. The pump wavelength is tuned between 1750 and 2050 nm, which results in a crossing point of the THRR and 5th harmonic emission a  pump wavelength of $\sim$ 1960 nm. SRR peak is generated at 390 nm. (b) THRR and 5th harmonic peak maxima vs pump wavelength, the points are measured along the solid red and black lines in (a). We find an enhancement of 40$\%$ compared to that predicted by taking the sum of the average values away from their overlap wavelength (dashed red and black lines). This suggests the generation of a stimulated SRR peak centred at 390 nm.}
\label{diamond}
\end{figure}

In Fig. \ref{diamond}(a) we show the high-energy part of the output spectrum after $L=500$ $\mu$m propagation, when progressively varying the input pulse wavelength from $1750$ nm to $2050$ nm. We can observe the 5th-harmonic peak shifting linearly towards longer wavelengths as expected [see red solid line and red squares in Fig. \ref{diamond}(a) that shows the pump wavelength $\times 1/5$]. However, an additional peak is observed, which shifts towards shorter wavelengths when increasing the pump wavelength. This latter peak is due to THRR as demonstrated by the very good quantitative agreement with the predicted THRR position using the experimental parameters  [black solid line and black dots in Fig. \ref{diamond}(a)]. The prediction is based on the THRR phase-matching condition in normal dispersion and in presence of the crucial shock term, as in Ref. \cite{tom}. When the THRR and 5th harmonic peaks have similar frequencies, i.e. when the pump wavelength is $\sim 1960$ nm, the THRR amplitude grows considerably. The limited propagation distance, combined with the normal dispersion of diamond, does not allow the formation of a soliton, i.e. the pulse intensity will quickly decrease in propagation: yet these results show that the phenomenon of SRR `promotion' is very general and relies only on the crossing of the THRR emission with a higher-order harmonic.

Figure \ref{diamond}(b) shows the peak intensities of the 5th harmonic and THRR taken along the red and black solid lines from Fig. 6(a) respectively. There is a clear enhancement of the peaks at the point at which their emission wavelengths are overlapped ($\lambda_{pump} \sim 1960$ nm). We find that there is a significant enhancement of the combined peak [$\sim$40$\%$ larger than predicted, see blue dashed line] indicating the production of a {\em stimulated} SRR peak.

\paragraph{Conclusions ---} 
In this paper we have shown that the THRR can be stimulated by a high-harmonic when they are close in the spectral domain. The numerical simulations show that resonant radiation peak could grow indefinitely in a stimulated fashion, with its amplitude even becoming {\em higher} than the pump pulse amplitude itself in some cases. This is a two step process, as suggested by the XFROG diagrams of our simulations. We have seen experimentally some preliminary hints of SRR in diamond, where a very intense pulse propagates in normal dispersion and the radiation is stimulated by the 5th harmonic. We expect this same phenomenon to appear in any nonlinear system in which the resonant radiation associated with a higher harmonic generation term in the evolution equation can be phase matched to a frequency in which it overlaps with the higher harmonic itself. For example, preliminary investigation shows the same effects to appear in the KdV equations governing the evolution of hydrodynamics and will be reported in future work. The discovery of this harmonic-mediated resonance could lead, by using appropriate waveguides or bulk crystals, to super-efficient frequency conversion effects.

\end{document}